\documentclass[aps,twocolumn,superscriptaddress,amsmath,amssymb,prl]{revtex4-2}
\usepackage{graphicx}
\usepackage{dcolumn}
\usepackage{bm}
\usepackage{braket}
\usepackage{appendix}
\usepackage{subfigure}
\usepackage{hyperref}
\usepackage{color}
\usepackage{xcolor}
\usepackage{amsmath}
\usepackage{nccmath}
\usepackage[english]{babel}
\usepackage{svg}

\begin{document}
\title{On-chip optical levitation with a metalens in vacuum}

\author{Kunhong Shen}
\affiliation{Department of Physics and Astronomy, Purdue University, West Lafayette, Indiana 47907, USA}

\author{Yao Duan}
\affiliation{Department of Electrical Engineering, The Pennsylvania State University, University Park, Pennsylvania 16802, USA}

\author{Peng Ju}
\affiliation{Department of Physics and Astronomy, Purdue University, West Lafayette, Indiana 47907, USA}

\author{Zhujing Xu}
\affiliation{Department of Physics and Astronomy, Purdue University, West Lafayette, Indiana 47907, USA}

\author{Xi Chen}
\affiliation{Department of Electrical Engineering, The Pennsylvania State University, University Park, Pennsylvania 16802, USA}

\author{Lidan Zhang}
\affiliation{Department of Electrical Engineering, The Pennsylvania State University, University Park, Pennsylvania 16802, USA}

\author{Jonghoon Ahn}
\affiliation{School of Electrical and Computer Engineering, Purdue University, West Lafayette, Indiana 47907, USA}

\author{Xingjie Ni}
\email{xingjie@psu.edu}
\affiliation{Department of Electrical Engineering, The Pennsylvania State University, University Park, Pennsylvania 16802, USA}

\author{Tongcang Li}
\email{tcli@purdue.edu}
\affiliation{Department of Physics and Astronomy, Purdue University, West Lafayette, Indiana 47907, USA}
\affiliation{School of Electrical and Computer Engineering, Purdue University, West Lafayette, Indiana 47907, USA}
\affiliation
{Birck Nanotechnology Center, Purdue University, West Lafayette,
	IN 47907, USA}
\affiliation
{Purdue Quantum Science and Engineering Institute, Purdue University, West Lafayette, Indiana 47907, USA}

\date{\today}

\begin{abstract}
Optical levitation of dielectric particles in vacuum is a powerful technique for precision measurements, testing fundamental physics, and quantum information science. Conventional optical tweezers  requires bulky optical components for trapping and detection. Here we design and fabricate an ultrathin dielectric metalens with a high numerical aperture of 0.88 at 1064 nm in vacuum. It consists of 500 nm-thick silicon nano-antennas, which are compatible with ultrahigh vacuum. We demonstrate optical levitation of nanoparticles in vacuum with a single metalens. The trapping frequency can be tuned by changing the laser power and polarization. We also transfer a levitated nanoparticle between two separated optical tweezers. Optical levitation with an ultrathin metalens in vacuum provides opportunities for a wide range of applications including on-chip sensing. Such metalenses will also be useful for trapping ultacold atoms and molecules.
\end{abstract}

\maketitle


There is remarkable progress in levitated optomechanics  over the past decade \cite{Millen_2020}, including ultrasensitive force and torque detection \cite{ranjit2016zeptonewton,Ahn2020}, acceleration sensing \cite{Monteiro2020PRAacclereation}, mass measurement \cite{Zheng2020robust}, GHz rotation \cite{PhysRevLett.121.033603,Reimann2018GHz,Ahn2020}, chemical nano-reactor \cite{ricci2021chemical}, optical refrigeration \cite{rahman2017laser}, quantum ground-state cooling \cite{Delic892,magrini2021realtime,Tebbenjohanns2021quantum}, and testing fundamental physics. A levitated dielectric particle in vacuum is one of the most promising systems for studying macroscopic quantum effects since it is well isolated from the thermal environment. Several schemes have been proposed for utilizing optical levitation systems to study Casimir physics \cite{XuJacobLi2021}, quantum aspects of gravity \cite{BosePRL2017}, and search for dark matter and dark energy \cite{rider2016search}. 
Typically, optical trapping of a nanoparticle in vacuum uses a bulky objective lens with a large numerical aperture (NA) to tightly focus a laser beam \cite{PhysRevLett.121.033603,Reimann2018GHz,Ahn2020}. If the NA of the focusing lens is not high enough, a pair of lenses would be required to focus two counter-propagating laser beams for creating  a stable three-dimensional (3D) trap without the help of gravity \cite{li2010measurement,ranjit2016zeptonewton}. The system is usually bulky and inconvenient for practical applications of optically levitated particles, such as accelerometers. Minimizing the focusing lens and levitating a dielectric particle on chip are on demand for realizing compact levitated optomechanical systems with high performance and efficiency. Here we report on-chip optical levitation of nanoparticles in vacuum with an ultrathin nanofabricated metalens.

\begin{figure*}[!ht]

  \centering
  \includegraphics[width=0.99\linewidth]{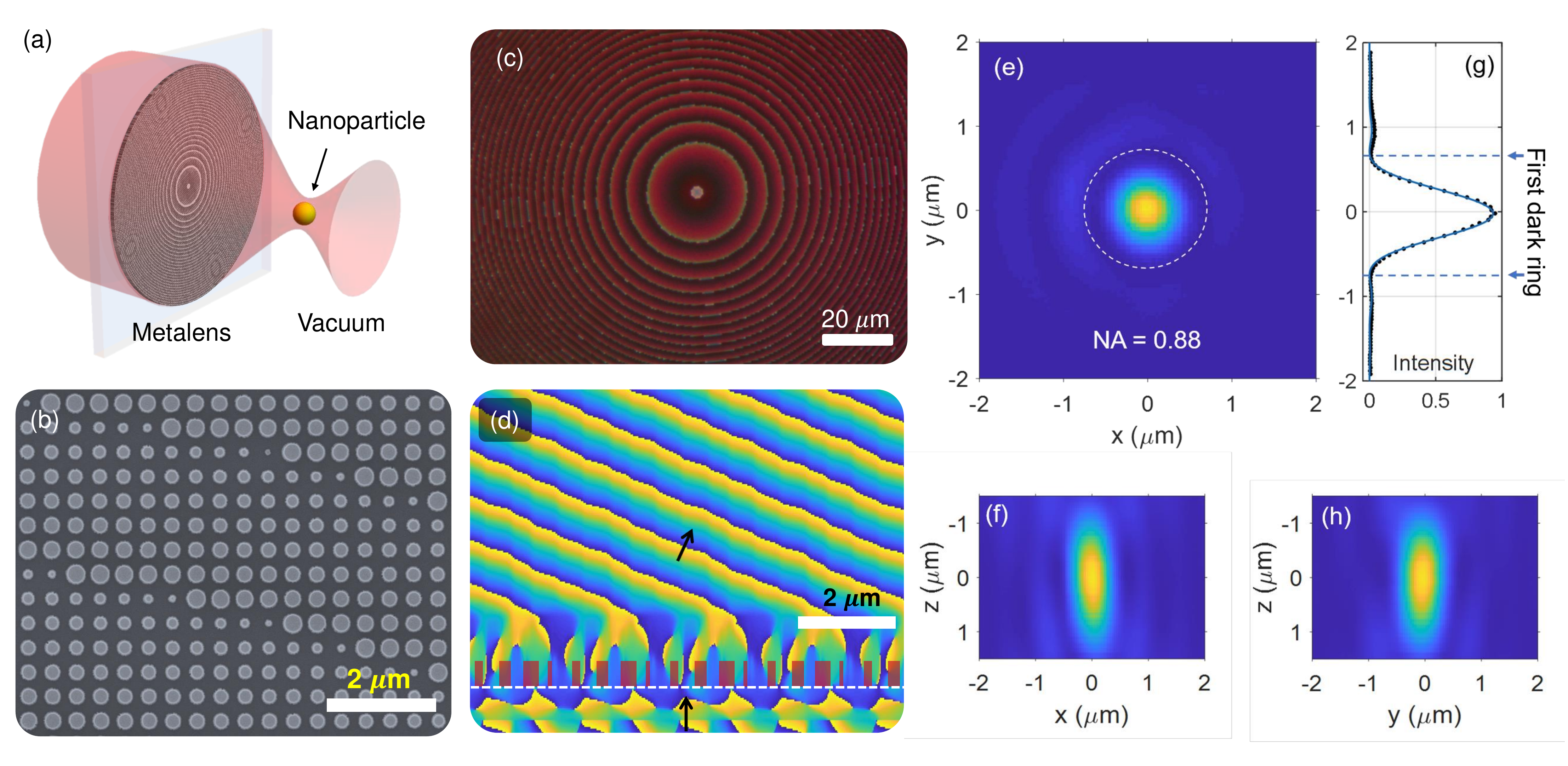}
  \caption{(a) Schematic of an optically levitated nanoparticle with a metalens in vacuum. (b) A SEM image of the metalens antennas, which are silicon nanopillars with a height of 500 nm. (c) An optical image of a metalens.  (d) One-dimensional phase simulation of a beam passing through the antennas. The antennas are shown as brown pillars and the substrate surface of the metalens is denoted by the white dashed line. (e), (f) and (h) are the measured laser profiles in $z=0$, $y=0$ and $x=0$ planes, respectively. The white dashed line in (e) indicts the first dark ring of the Airy's disk. (g) The laser intensity  along the $y$ axis ($x=z=0$). Dots are the experimental data. The solid line is an Airy function for the diffraction by a circular aperture. The radius of the first dark ring is 740 $\mu$m.}
  \label{fig:1}

\end{figure*}

An optical metasurface is a recent breakthrough in optics \cite{yu2011light}. It uses nanostructures with features typically smaller than the  wavelength   to achieve unprecedented control of light properties \cite{ni2015ultrathin}. A metalens may be created by arranging phase-shifting elements on a surface that forms a spherical phase profile to function as a spherical lens \cite{khorasaninejad2016metalenses}. 
Optical trapping in liquids based metalenses were demonstrated  recently \cite{doi:10.1021/acs.nanolett.8b01844,Tkachenko:18,Plidschun2021}. 
 Optical manipulation and trapping by metalens show high flexibility for being integrated to a chip-based device and fluid cells. 
However,  metalens optical traps have only been carried out in liquids so far \cite{doi:10.1021/acs.nanolett.8b01844,Tkachenko:18,Plidschun2021}, which  limits the scope of applications. The largest full angular aperture of the metalenses used in these trapping experiments is 85$^\circ$,  corresponding to an NA of 0.9 in water \cite{Plidschun2021}.  Because of the larger contrast of the refractive index and hence more light scattering in vacuum,  single-beam optical trapping of a nanoparticle in vacuum requires  a focusing lens with  a larger  angular aperture than that in a liquid. In addition, the effect of laser heating is more severe  in vacuum.
To the best of our knowledge, there has been no report of on optical levitation of particles with a metalens in air or vacuum. 

\begin{figure*}[!ht]
  \centering
  \includegraphics[width=0.99\linewidth]{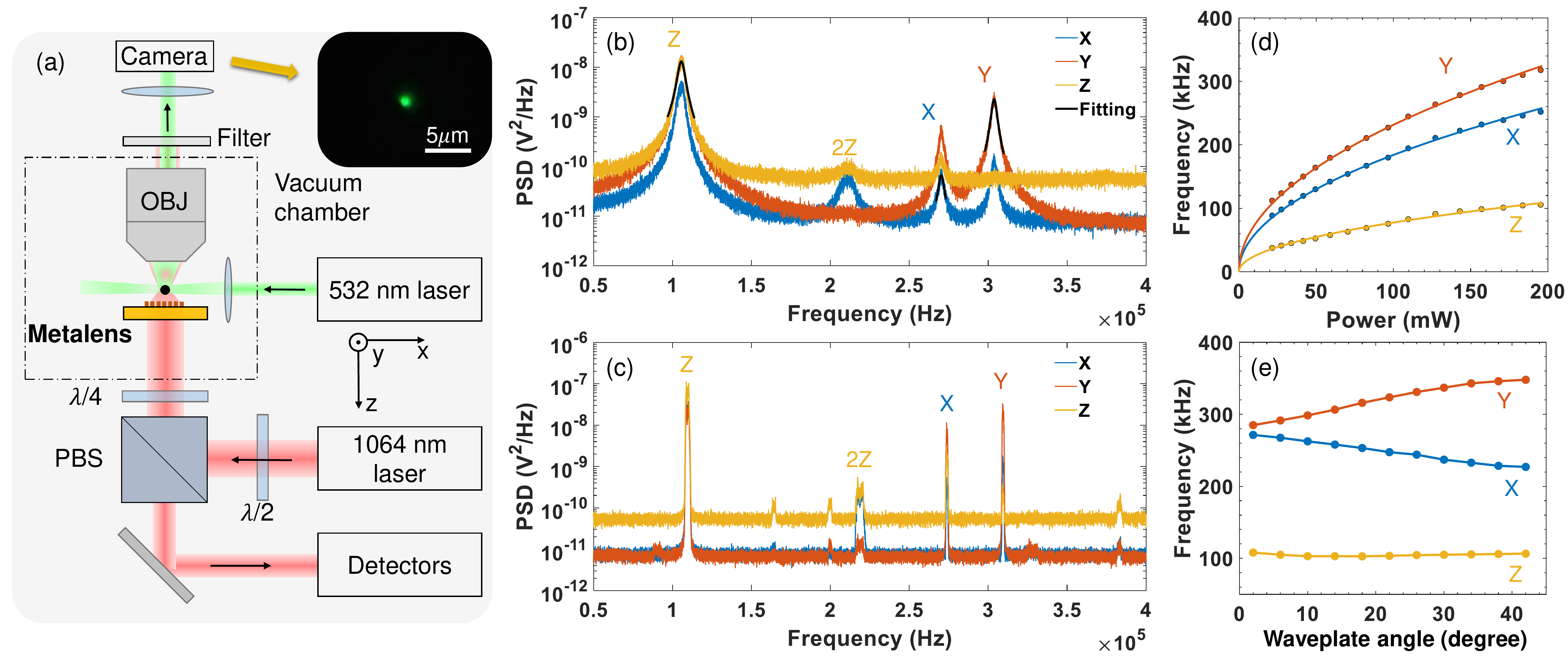}
  \caption{(a) Schematics of the metalens-based optical levitation and detection system. A 1064-nm laser is used for optical levitation and a 532-nm laser is used for imaging.  The inset figure is the optical image of a levitated particle. PBS: polarizing beamsplitter. OBJ: objective lens. (b) and (c) Power spectral density (PSD) of the mechanical motions of the levitated particle at 5 Torr and 5 mTorr with an elliptically polarized trapping laser.  The black solid curves in (b) are the Lorentzian fittings. The  peak around 215 kHz is the second-harmonic generation of z peak. Other small peaks in (c) come from difference-frequency generations and sum-frequency generations. (d) The relationship between the trapping frequencies is shown as a function of the laser power. Solid lines are the fitting with a square root function.  (e) The measured trapping frequencies are shown as a function of the laser polarization.}
  \label{fig:2}
\end{figure*}

In this letter, we report the  realization of on-chip optical levitation in vacuum with a  high-NA metalens (Fig. \ref{fig:1}). We design and fabricate planar metalenses with silicon nanopillars on a sapphire substrate. The full angular aperture of our metalens is 123$^\circ$, corresponding to a measured NA of  0.88 in air. We levitate a nanoparticle by optical tweezers generated with a single metalens at a pressure of $2\times 10^{-4}$ Torr without feedback cooling. We then transfer a nanoparticle between two  optical tweezers separated by about 1.5 $\mu$m.  Our work demonstrates the feasibility of using nanofabricated metalens in nanoparticle trapping in vacuum. Metalenses for multiple trapping wells and more complex potentials  can be fabricated  in the future.


We first design and fabricate a high-NA metalens for levitating a nanoparticle in vacuum (Fig. \ref{fig:1}(a)). 
 Assuming the incident laser propagates along the $z$-axis, a high-NA lens should have an aspheric phase shift $\phi(x,y)=2\pi/\lambda\cdot(f-\sqrt{x^2+y^2+f^2})+\phi_0$, where $\lambda$ is the wavelength, $f$ is the focal length, and $\phi_0$ is an overall phase constant. The parabolic phase can be shifted  by integer numbers of $2\pi$  to reduce the phase range to $2\pi$.  The metalens is designed to have a diameter of 425 $\mu$m, a focal length of 100 $\mu$m, and a numerical aperture of 0.9  for a 1064 nm laser beam in vacuum. 
 We create the metalens using amorphous silicon nanopillars (Fig. \ref{fig:1}(b)), which have a large refractive index of 3.6 at 1064 nm.  The silicon nanopillars work as antennas for the laser beam.   The phase shift is controlled by the different diameters of the silicon antenna. The height of silicon nanopillars is 500 nm.  Fig. \ref{fig:1}(d) reveals the simulated phase variation of the electric field propagation on both sides of the metalens.  We deposit 500 nm thick amorphous silicon on a  sapphire substrate in plasma-enhanced chemical vapor deposition (PECVD) tool under 220 $^\circ$C, and create the nanopillars with e-beam lithography.      These materials are compatible with ultrahigh vacuum. 
 Figure \ref{fig:1}(b) shows a SEM image of a metalens which consists of multiple nanometer scale silicon antennas. Fig.  \ref{fig:1}(c) shows an optical image of the metalens.


The laser profiles at the focal spot along different directions are measured to determine the NA of the metalens. To measure the NA of the metalens, a 1064 nm laser with a beam waist much larger than the diameter of the metalens is incident on the metalens. So it is convenient to consider the incident beam as a plane wave with a constant amplitude. We use a conventional NA=0.9 objective lens to collect the focused laser beam after the metalens, and take images with a camera. The metalens is moved along $z$-axis with a motorized actuator to measure the laser profile along $z$-axis.
 Fig. \ref{fig:1}(e),(f) and (h) present the measured laser profiles at the focal point in $z=0$, $y=0$ and $x=0$  planes, respectively.
 By fitting the first dark ring of the Airy's disc, the NA of the metalens is measured to be 0.88, which is close to the designed value.  


Now we demonstrate the first observation of metalens-based optical levitation in vacuum. The experimental setup is shown in Fig. \ref{fig:2}(a).
The metalens is mounted on a translation stage. A 1064-nm laser beam is directed onto the metalens for optical levitation. A 532-nm laser illuminates the levitated nanoparticle for imaging with a conventional objective lens (OBJ).  The Gaussian beam diameter of the 1064-nm laser is adjusted to 425 $\mu$m to match  the diameter of the metalens, which improves the levitation stability and  the trapping efficiency. The incident laser power at focus is around 200 mW. 

\begin{figure}[thb]
  \centering
  \includegraphics[width=0.99\linewidth]{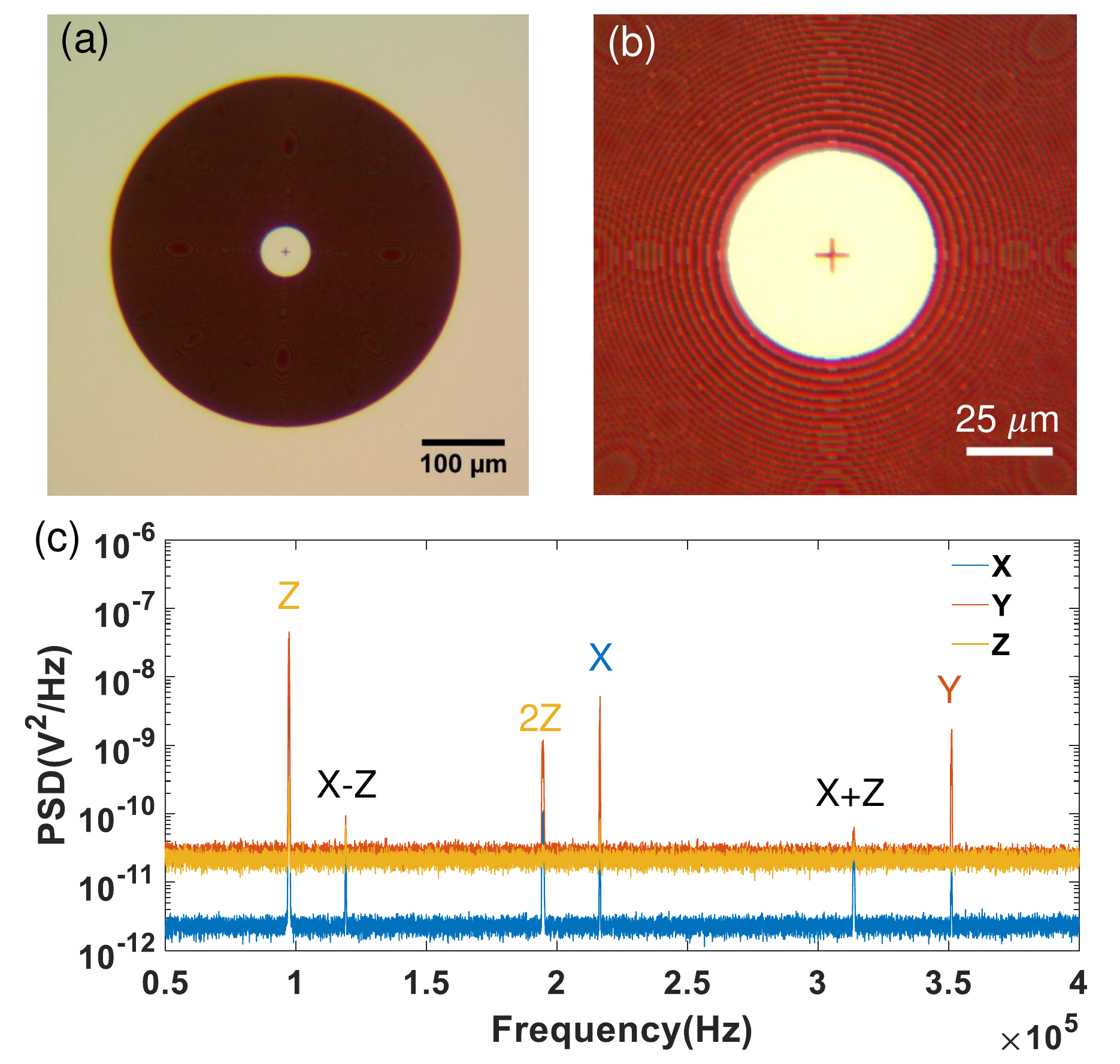}
  \caption{(a), (b) Otical images of a metalens with a 60-$\mu$m-diameter hole at the center. The center cross mark is used for alignment. The focal length of the metalens is 100 $\mu$m. (c) PSD of a nanoparticle at $2.0\times10^{-4}$ Torr with a linearly polarized incident beam. The trapping frequencies of x, y and z motions are about 216, 351 and 97 kHz.}
  \label{fig:3}
\end{figure}

In the experiment, silica nanoparticles with a diameter of 170~nm are used.
The nanoparticles are diluted in  deionized water and launched by an ultrasonic nebulizer  for trapping. The metalens can successfully levitate nanoparticles even after repeated use.
The inset of Fig. \ref{fig:2}.(a) shows an optical image of a  trapped nanoparticle using the scattered light of the 532-nm laser.
To measure the motion of a trapped nanoparticle, a set of  balanced detectors collect the  light backward scattered by the nanoparticle and the reflected light from the surface of the sapphire substrate of the metalens. The scattered light from the trapped particle interferes with the reflection light from the substrate and hence we can measure the motion of the particles.
Compared to the typical  backward detection which requires an additional independent reference laser beam for interference, the detection with the metalens system is much simpler and  can be integrated to chip-based devices.

\begin{figure}[ht]
    \centering
    \includegraphics[width=0.99\linewidth]{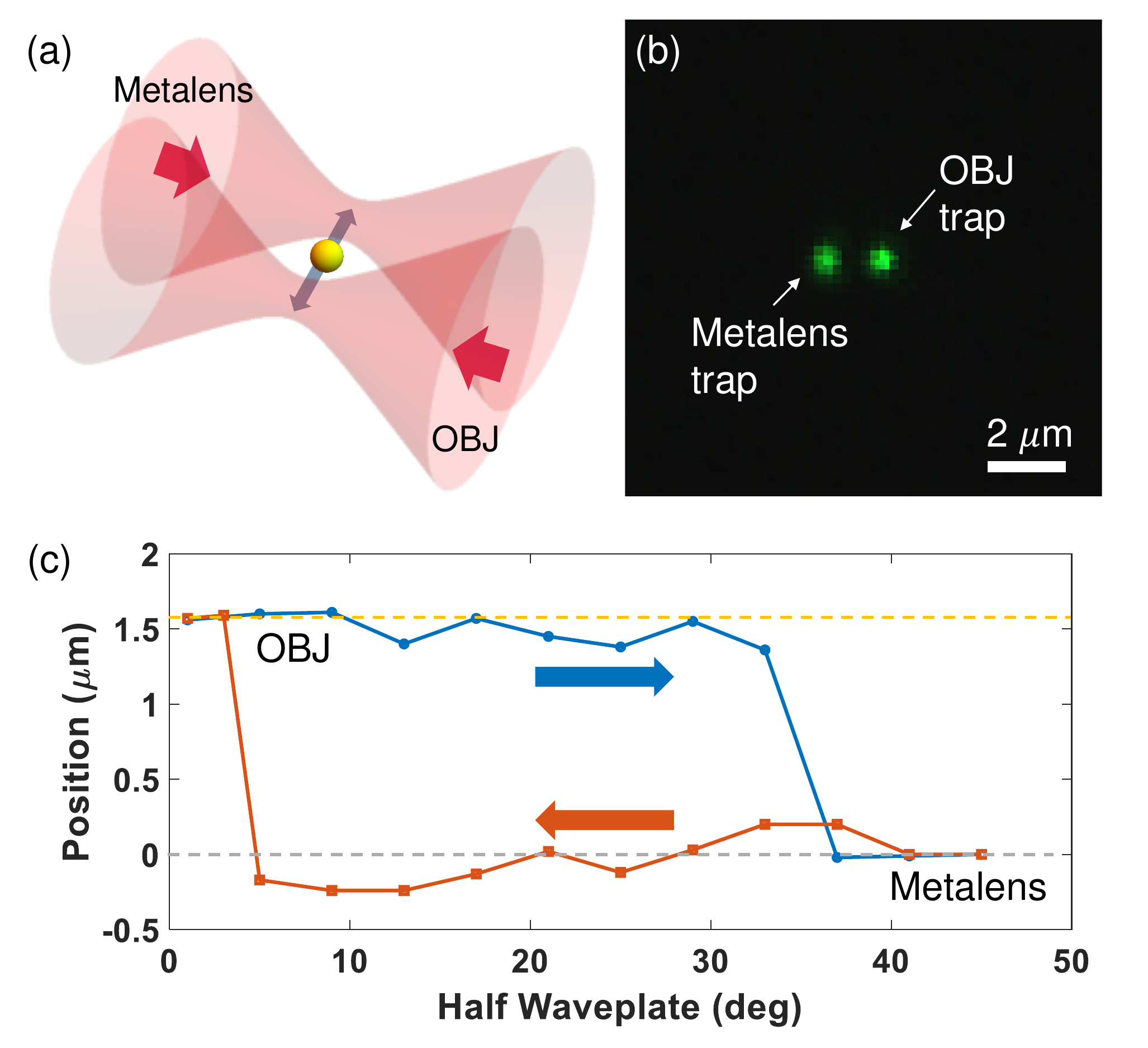}
    \caption{(a) Scheme of a nanoparticle levitated by a dual-beam trap. (b) Optical image of a particle transferring between two potential wells. The left (right) green spot shows the particle was captured by the metalens (OBJ). The separation between two wells is about 1.5 $\mu$m. (c) The transfer process was adjusted by controlling the laser power distribution. Zero degree (45$^\circ$) of the half waveplate means the total laser power was directed on OBJ (metalens). The dashed yellow (gray) line is the steady trapping position for OBJ (metalens) trap.}
    \label{fig:4}
\end{figure}

In the frequency domain, the power spectrum density (PSD) of the motion of a levitated nanoparticle is given by \cite{li2011millikelvin}
\begin{equation}
    S_{qq}(\omega)=\frac{\Gamma_{\mathrm{CM}}k_\mathrm{B} T_{\mathrm{CM}}/\pi M}{(\omega^2-\omega_\mathrm{q}^2)^2+\Gamma_{\mathrm{CM}}^2\omega^2},\notag
\end{equation}
where $\Gamma_\mathrm{CM}$ is the damping rate of the center-of-mass motion (COM), $\omega_\mathrm{q}$ is the mechanical oscillation frequency of the trapped particle, $T_\mathrm{CM}$ is the temperature of center-of-mass motion, M is the mass of the particle and $k_\mathrm{B}$ is the Boltzmann constant. 
Fig. \ref{fig:2}(b) and (c) are the PSD of  the center-of-mass motion at the pressure of 5 Torr and 5 mTorr, respectively. The black solid curves are the Lorentzian fittings. The trapping frequencies for $x$,  $y$, and $z$ directions are  270 kHz, 304 kHz, and  106 kHz respectively.  They are  higher than the frequencies achieved with a conventional NA=0.9 objective lens at the same trapping laser power.
Fig. \ref{fig:2}(d) shows that the trapping frequencies are proportional to the square root of laser power, which is similar to the case that the particle is trapped by OBJ. 
Meanwhile, the trapping frequency is also related to the laser polarization. If the incident beam is linearly polarized, its trapping frequency in $y$ direction can reach up to 350 kHz as shown in Fig.  \ref{fig:2}(e). The higher trapping frequency will be helpful for ground-state cooling. 
In addition, PSD in Fig. \ref{fig:2}(b) and(c) shows that the trapping potential well is not perfectly harmonic. Strong nonlinear effects is observed including second-harmonic, sum-frequency, and difference-frequency generation.

The transmitted beam from the  center part of a lens contributes mainly to the scattering force. We may eliminate the central part of the beam without hurting  optical levitation. 
 In Fig. \ref{fig:3}(a) and (b), we show a metalens with a hole at the center. The hole diameter is  60 $\mu$m. 
Experimentally, we find that the metalenses with  a 60-$\mu$m hole has a  better performance. It can trap a nanoparticle at a lower pressure than a metalens without a hole when there is no feedback cooling. Fig. \ref{fig:3}(c) is the PSD of a nanoparticle trapped at $2.0\times10^{-4}$ Torr with a linearly polarized beam by metalens with a 60-$\mu$m hole. 


We also realize optical levitation in a dual-beam trap and transfer a nanoparticle between two optical tweezers. One beam is focused by the metalens. The other trapping beam is implemented by a conventional OBJ, as shown in Fig. \ref{fig:4}(a).
The particle in the dual-beam trap can jump from one potential well to the other, depending on the laser power distribution. An optical image of the particle in a dual-beam trap is shown in \ref{fig:4}(b).
The separation between two potential wells is about 1.5 $\mu$m. In Fig. \ref{fig:4}(c), we show the transfer process between two trapping wells by controlling laser power distribution. When the half waveplate is at 0$^\circ$  (45$^\circ$ ), all of the laser power is directed on the OBJ lens (metalens). Because the maximum trapping depth is  large, the particle jumping happens when the laser power of one optical tweezers is close to zero. The transfer process is reversible and repeatable.


In conclusion, we have demonstrated the first experimental realization of metalens-based on-chip optical levitation of a nanoparticle  in vacuum. 
With a high NA metalens, we  achieve high trapping frequencies. 
Compared to a conventional objective lens, our metalens can work at more extreme conditions. It also provides more freedom to generate complex trapping potentials with nanofabrication.  Metalens levitation in vacuum can open more opportunities to study fundamental physics and applied science. For example, we can precisely control the distance between a surface and a trapped particle for studying surface interactions \cite{XuJacobLi2021}.
In addition, a high-NA metalens  can potentially replace specially-designed bulky objective lenses  \cite{robens2017high} to create tightly-focused optical tweezers for trapping ultracold atoms and molecules in vacuum for quantum simulation \cite{bernien2017probing} and ultracold chemistry \cite{anderegg2019optical}.

We acknowledge supports from the Office of Naval Research (ONR) Basic Research Challenge (BRC) program (N00014-18-1-2371), and the Laboratory Directed Research and Development program at Sandia National Laboratories.
We thank  Alejandro J. Grine and Francis Robicheaux for helpful discussions.

\end{document}